# Possible high-$T_c$ superconductivity at 45 K in the Ge-doped cluster Mott insulator GaNb$_4$Se$_8$

Ji-Hai Yuan[1,2,#], Ya-Dong Gu[1,2,#], Yun-Qing Shi[1,2], Hao-Yu He[1,2], Qing-Song Liu[1,2], Jun-Kun Yi[1,2], Le-Wei Chen[1,2], Zheng-Xin Lin[1,2], Jia-Sheng Liu[1,2], Meng Wang[1,2], Zhi-An Ren[1,2,*]

[1] Institute of Physics and Beijing National Laboratory for Condensed Matter Physics, Chinese Academy of Sciences, Beijing 100190, China

[2] School of Physical Sciences, University of Chinese Academy of Sciences, Beijing 100049, China

[#] These authors contributed equally to this work.

[*] Corresponding author. E-mail: renzhian@iphy.ac.cn

**Abstract**

Polycrystalline samples of Ge-doped GaNb$_4$Se$_8$ were synthesized via the conventional solid-state reaction method. One batch of samples among our synthesis attempts exhibited clear zero resistance transitions, with the highest onset superconducting critical temperature $T_c$ reaching 45 K and full zero resistance achieved near 25 K. This observation provides the first tentative evidence of superconductivity induced by electron doping in the cluster compound GaNb$_4$Se$_8$, and it may demonstrate a new class of Nb-based high-$T_c$ superconductors arising from doped Mott insulators upon further magnetic susceptibility confirmation.

In 1986, Bednorz and Müller discovered superconductivity in the Ba-doped $La_2CuO_4$ with a critical temperature ($T_c$) about 35 K, initiating a new era for high-$T_c$ superconductivity researches [1]. $La_2CuO_4$ belongs to a broad family of layered copper oxides which are characterized by strongly correlated electrons localized within the two-dimensional $CuO_2$ planes, and these materials exhibit antiferromagnetic long-range order and behave as Mott insulators. The introduction of charge carriers by chemical doping into the $CuO_2$ planes, either by holes or electrons, can break the electronic correlations and drive the system through a rich sequence of electronic phases, including strange metal, pseudogap, stripe phase, and high-$T_c$ superconductivity, *etc*. Building upon this paradigm, the pioneering work was rapidly surpassed by the $T_c$ being raised to 93 K in $YBa_2Cu_3O_7$ above the boiling point of liquid nitrogen (77 K) and a record of 135 K in $HgBa_2Ca_2Cu_3O_8$ that remains unbroken [2-3]. These achievements underscored the central role of electronic correlations and quasi-two-dimensional crystal structures in enabling unconventional pairing mechanisms of high-$T_c$ cuprate superconductors. The recent discovery of iron-based and nickel-based superconductors in this century has further demonstrated that tuning electronic correlations through carrier doping in strongly correlated electron systems is a pivotal route toward the discovery of novel high-$T_c$ superconductors [4-10].

$GaNb_4Se_8$ is a well-known cubic lacunar spinel which holds a 148-type general formula $AM_4X_8$ (A = Al, Ga, Ge; M = V, Nb, Mo, Ta; X = chalcogen). It was first synthesized by Benyaich *et al*. as a cluster compound [11], in which $Nb_4$ tetramers are formed in a pyrochlore-type network as shown in Fig. 1. In $GaNb_4Se_8$, an unpaired electron occupies the molecular $t_2$ orbital, and forms a nonmagnetic Mott insulating ground state. The interplay among electron correlations, spin-orbit coupling and Jahn-Teller effect gives rise to complex physics in $GaNb_4Se_8$. At $T_Q$ = 50 K, $GaNb_4Se_8$ undergoes a cubic-to-cubic structural transition accompanied by quadrupolar ordering of molecular orbitals on $Nb_4$ clusters. Upon further cooling to $T_M$ = 31 K, anisotropic lattice distortions emerge, transforming the crystal structure into an orthorhombic lattice and inducing local rotations of the $Nb_4$ clusters [12-18]. By applying high pressure, $GaNb_4Se_8$ changes from Mott insulator into a metallic and superconducting state with a $T_c$ of 2.9 K under 13 GPa [19]. However, no successful carrier doping study has been reported on $GaNb_4Se_8$ to date.

To investigate the electron doping effects in GaNb$_4$Se$_8$, we synthesized Ge-substituted samples (Ga$_{1-x}$Ge$_x$Nb$_4$Se$_8$) via chemical substitution at the Ga sites and studied their electrical transport properties. This substitution is expected to introduce an extra electron into the molecular $t_2$ orbital of the Nb$_4$ clusters, thereby destabilizing the Mott insulating state and potentially driving the system toward a metallic phase (Fig. 1). Here we report the possible high-$T_c$ superconductivity observed in Ge-doped GaNb$_4$Se$_8$ evidenced by explicit zero resistance transitions.

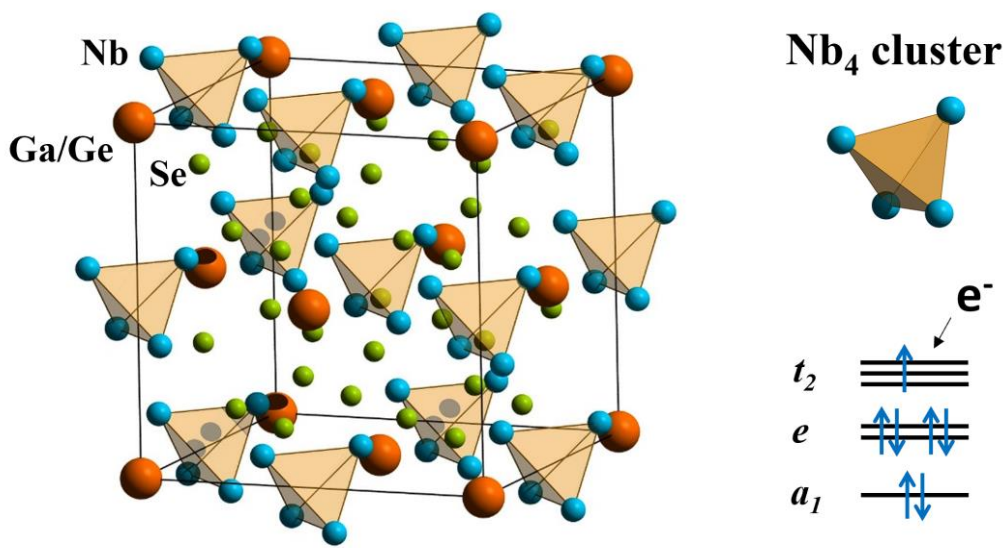

Figure 1. Schematic crystal structure of Ge-doped GaNb$_4$Se$_8$, the Nb$_4$ tetrahedron cluster, and the molecular orbital scheme of the Nb$_4$ tetramer.

The Ge-doped Ga$_{1-x}$Ge$_x$Nb$_4$Se$_8$ polycrystalline samples were synthesized by conventional solid-state reaction method. High-purity (99.99%) Ga, Nb, Se, Ge ingredients were weighed and sealed in an evacuated quartz ampule. The ampule was heated at 900 °C for 72 hours. The resulting mixture was reground thoroughly and pressed into pellets, which were sealed in an evacuated quartz ampule again and heated at 1000 °C for another 72 hours. All the preparation procedures were carried out in a glove box filled with high-purity Ar gas to avoid any possible contamination. The obtained samples are black in color and seem stable in air. We note that GeSe crystals are often observed accumulating at the end of the quartz ampule due to its volatility at lower temperatures, which makes the doping of Ge at Ga site extremely difficult and uncontrollable. Powder X-ray diffraction (XRD) data were collected using a PAN-analytical X-ray diffractometer with Cu-K$_\alpha$ radiation. The electrical resistance measurements were performed on a Quantum Design physical property measurement system (PPMS) with the standard four-probe method. The *dc* magnetization was measured with a Quantum Design magnetic property measurement system.

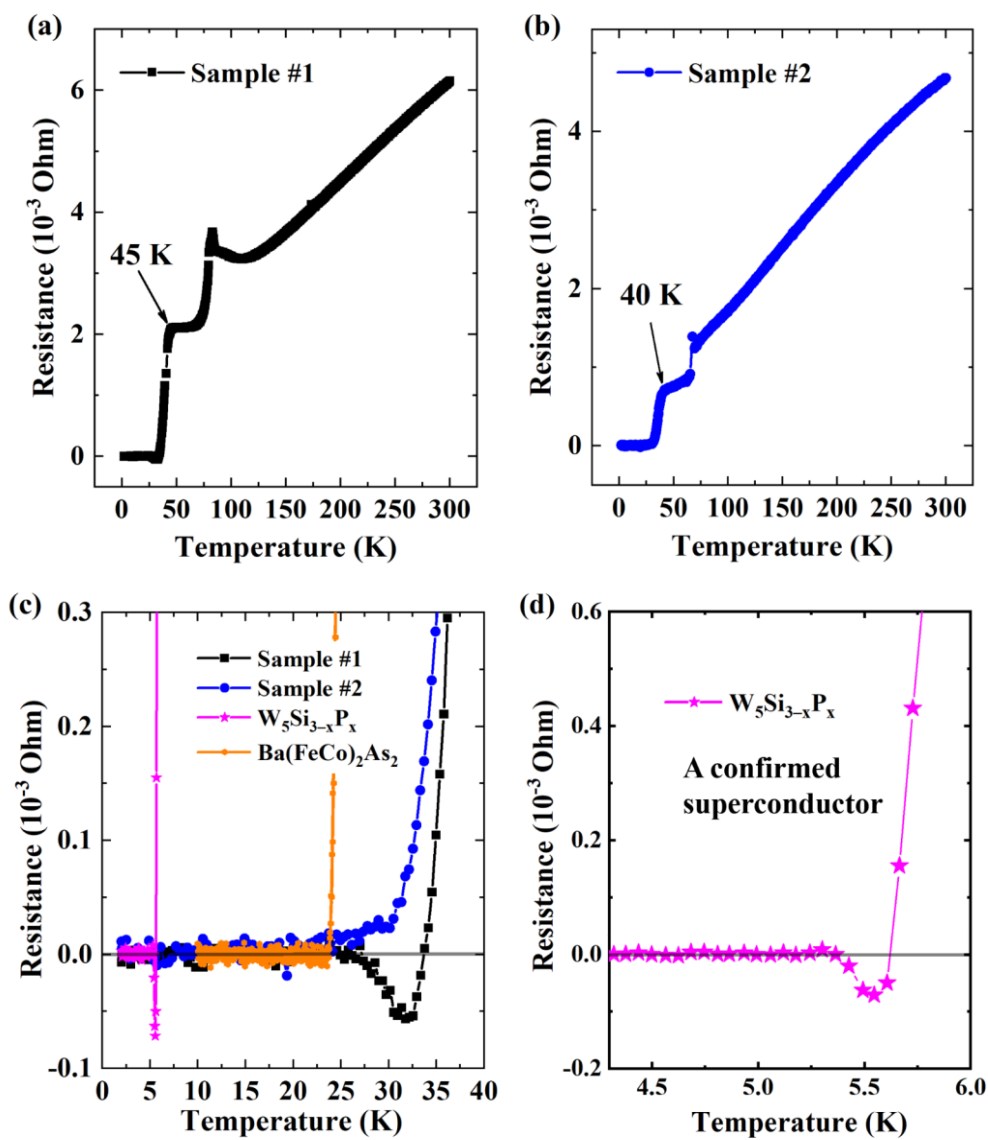

Figure 2. (a, b) Temperature dependence of the electrical resistance of Ge-doped GaNb$_4$Se$_8$ for two samples. (c) The zero resistance comparison with two other superconductors W$_5$Si$_{3-x}$P$_x$ and Ba(FeCo)$_2$As$_2$. (d) The similar negative resistance phenomenon observed in a W$_5$Si$_{3-x}$P$_x$ superconductor.

This study aims to investigate the effects of electron doping in the Mott insulator $GaNb_4Se_8$ through chemical substitution of Ge at Ga sites. We emphasize that, across hundreds of samples tested in our experiments, only a single batch of samples prepared within the same ampule shows zero resistance transitions around 40 K. The high volatility of GeSe within the synthesis temperature range poses significant challenges to the precise control of the chemical substitution for Ge at Ga sites. The nominal composition of this batch of samples is $Ga_{0.9}Ge_{0.2}Nb_4Se_8$, in which the excess Ge was intentionally added to compensate for the volatilizing loss, and we have also tried many other compositional combinations but without reproducible results.

Fig. 2 (a, b) shows the temperature dependence of resistance for sample #1 and #2 with the applied current of 1 mA. We note that for these resistance measurements, the samples have irregular shapes as they were cut without any polishing, and the size is about 5 mm in length with a cross-sectional area about 1-2 $mm^2$. The doped samples exhibit metallic behaviors. Sample #1 and #2 display superconducting transitions with onset $T_c$ at 45 K and 40 K respectively, with zero resistance achieved at lower temperatures around 25 K. At higher temperatures around 60 - 80 K, both of the two samples show resistivity anomaly which may indicate another phase transition.

The zero resistance transitions of sample #1 and #2 are magnified and shown in Fig. 2c, and are compared with two other confirmed superconductors $W_5Si_{3-x}P_x$ and $Ba(FeCo)_2As_2$, which demonstrates good zero resistance superconductivity of both two samples (the zero resistance values correspond to measured voltage about $10^{-8}$ V, which approaches the lowest detection limit of PPMS). It's worth mentioning that below 34 K for sample #1, its resistance has a small segment of negative values. This phenomenon happens occasionally in inhomogeneous superconductors, and it is due to the reversal of the current direction between the two voltage probes in four-probe method, which originate from the dispersive superconducting percolation paths inside the sample when the superconducting channel is partly established. It disappears when the superconducting channel between two current probes is fully established at lower temperature, then the resistance drops to zero. A confirmed $W_5Si_{3-x}P_x$ superconductor in our previous studies exhibits the same negative resistance phenomenon at 5.5 K as shown in Fig. 2d [20]. We note that for this batch of samples, all superconducting signals disappeared after being stored in a glovebox for several days and the samples became insulating, and the zero resistance transitions are all the data that we have obtained to prove superconductivity. One explanation for the loss of superconducting signal is that the Ge-doped samples are unstable and gradually decompose into the parent compound GaNb$_4$Se$_8$ over time.

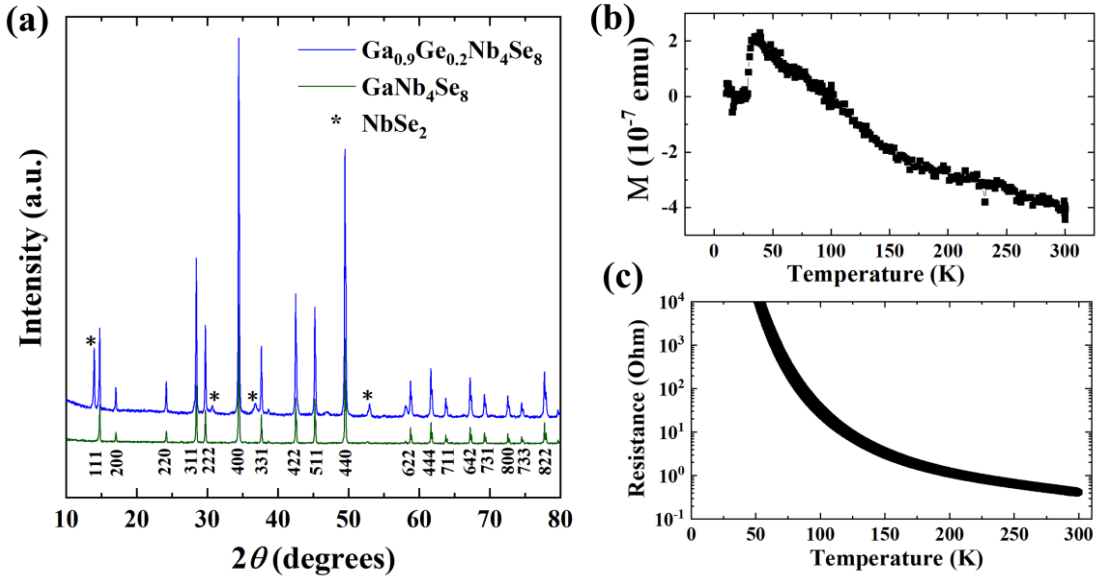

Figure 3. (a) Powder XRD patterns of the non-superconducting Ga$_{0.9}$Ge$_{0.2}$Nb$_4$Se$_8$ sample and undoped GaNb$_4$Se$_8$. (b) Temperature dependence of magnetization for the non-superconducting Ga$_{0.9}$Ge$_{0.2}$Nb$_4$Se$_8$ sample under 100 Oe. (c) Temperature dependence of electrical resistance for undoped GaNb$_4$Se$_8$.

After the loss of superconducting signals, the chemical phases of these samples were checked by powder XRD at ambient temperature, which is shown in Fig. 3a together with the data of a pure undoped GaNb$_4$Se$_8$ for comparison. NbSe$_2$ phase and other minor impurity peaks are observed in the Ga$_{0.9}$Ge$_{0.2}$Nb$_4$Se$_8$ sample. The main phase of GaNb$_4$Se$_8$ is nearly identical to its undoped phase in terms of lattice constant. In Fig. 3b, after the loss of zero resistance, the temperature dependence of magnetization for Ga$_{0.9}$Ge$_{0.2}$Nb$_4$Se$_8$ sample under a field of 100 Oe is exhibited, which is similar to that of undoped non-magnetic GaNb$_4$Se$_8$ in Ref. [14]. We note that the small drop of magnetization at 31 K originates from the structural transition in GaNb$_4$Se$_8$, and the negative values is caused by the background signal. Fig. 3c displays the temperature dependence of resistance for the undoped GaNb$_4$Se$_8$ which shows the typical Mott insulating behavior. For the clear zero resistance observed in the as-synthesized Ga$_{0.9}$Ge$_{0.2}$Nb$_4$Se$_8$ samples, the signal is more likely to originate from the main phase rather than from impurity phases. Nevertheless, further confirmation through magnetic susceptibility measurements remains essential to unambiguously establish the presence of superconductivity.

In summary, zero resistance was observed in a batch of Ge-doped cluster Mott insulator GaNb$_4$Se$_8$ samples with a superconducting-like transition onset $T_c$ at 45 K. This observation marks the first report of possible superconductivity emerging upon electron doping in this class of materials. Given the unique electronic structure involving correlated Nb$_4$ molecular orbitals and strong spin-orbit coupling, our findings suggest that GaNb$_4$Se$_8$ and related cluster Mott insulator compounds may constitute a promising new family of Nb-based high-$T_c$ superconductors.

**Acknowledgments**

This work was supported by the CAS Superconducting Research Project (Grant No. SCZX-0103), the National Key Research and Development Program of China (Grant No. 2021YFA1401800), the Strategic Priority Research Program of Chinese Academy of Sciences (Grant No. XDB25000000).